# Universality in long-distance geometry and quantum complexity

Adam R. Brown[1,2✉], Michael H. Freedman[3], Henry W. Lin[1,2,4] & Leonard Susskind[1,2]

In physics, two systems that radically differ at short scales can exhibit strikingly similar macroscopic behaviour: they are part of the same long-distance universality class[1]. Here we apply this viewpoint to geometry and initiate a program of classifying homogeneous metrics on group manifolds[2] by their long-distance properties. We show that many metrics on low-dimensional Lie groups have markedly different short-distance properties but nearly identical distance functions at long distances, and provide evidence that this phenomenon is even more robust in high dimensions. An application of these ideas of particular interest to physics and computer science is complexity geometry[3–7]—the study of quantum computational complexity using Riemannian geometry. We argue for the existence of a large universality class of definitions of quantum complexity, each linearly related to the other, a much finer-grained equivalence than typically considered. We conjecture that a new effective metric emerges at larger complexities that describes a broad class of complexity geometries, insensitive to various choices of microscopic penalty factors. We discuss the implications for recent conjectures in quantum gravity.

Universality is an idea that permeates physics and computer science. In various guises, this principle says that there are broad equivalence classes of phenomena at long distances, at long times or at high complexity that may be insensitive to the details of how processes are defined at short distances, short times or low complexity. This means that large, slow and highly complex experimentalists can make predictions that are robust against having to know the exact details of how everything works at the fundamental level.

In computer science, universality is integrated into the foundations of the field via the Church–Turing thesis. According to this thesis, and its associated theorems, a broad class of ways to define a computer all have the same computational power—that is, any function that can be computed in finite time by one machine in the class can be computed in finite time by any other member of the class. They all define the same set of computable functions. Universality is also foundational to complexity theory. To define the classes of functions that can be computed in polynomial time on a classical computer (P) or on a quantum computer (BQP), we do not need to be too careful about exactly which fundamental operations are permitted for our computer because a broad class of definitions are all equivalent. If there exist compilers such that two programming languages can each emulate the other with only polynomial overhead, then the set of functions they can compute in polynomial time is the same.

Universality also has a key role in physics. A large number of physical theories that differ in their predictions at the highest energies and shortest distance scales all give rise to approximately the same predictions at low energies and long distances. For example, in statistical physics, the Landau–Ginzburg theory of second-order phase transitions[8] says that the critical exponents generically depend only on the symmetries and are independent of any other details of the molecular dynamics. Similarly, in quantum field theory, almost all perturbations that can be made to the Hamiltonian at short distances (in the 'ultraviolet' (UV)) turn out to be irrelevant—that is, they have an ever smaller effect as we probe the system on longer and longer length scales (in the 'infrared' (IR)). The renormalized couplings associated with these irrelevant perturbations are said to decrease with length scale.

In this paper, we explore a kind of universality in long-distance geometry. We study the distance function on several curved spaces and argue that the distance between two well-separated points can be insensitive to most changes in the metric at short separations. We illustrate this with various examples and provide evidence that it is a generic feature of high-dimensional Riemannian geometries.

We then connect this phenomenon in differential geometry back to complexity theory. The bridge is complexity geometry[3–7] (Methods), which defines the complexity of performing operations on a quantum computer as a distance function on a high-dimensional group manifold. Through this definition, the conjectured universality of geometries becomes a universality of definitions of quantum complexity. Unlike the universality of computability, which allows any computational overhead as long as it is finite, or the universality of BQP, which demands only polynomial equivalence between runtimes, our conjectured universality allows at most linear overhead. We argue that this much more fine-grained universality is nevertheless robust. Finally, we describe how our results support previous conjectures about the connection between computational complexity and the properties of holographic black holes in quantum gravity.

## Motivating example

Let us motivate our program with an example. Consider a homogeneous three-dimensional space, of finite volume and a diameter of one metre. Define a 'small' deformation of this metric as one that preserves homogeneity and does not change the distance between any pair of points by more than one picometre.

[1]Google DeepMind, Mountain View, CA, USA. [2]Department of Physics, Stanford University, Stanford, CA, USA. [3]Department of Mathematics, University of California, Santa Barbara, Santa Barbara, CA, USA. [4]Department of Physics, Princeton University, Princeton, NJ, USA. ✉e-mail: mr.adam.brown@gmail.com

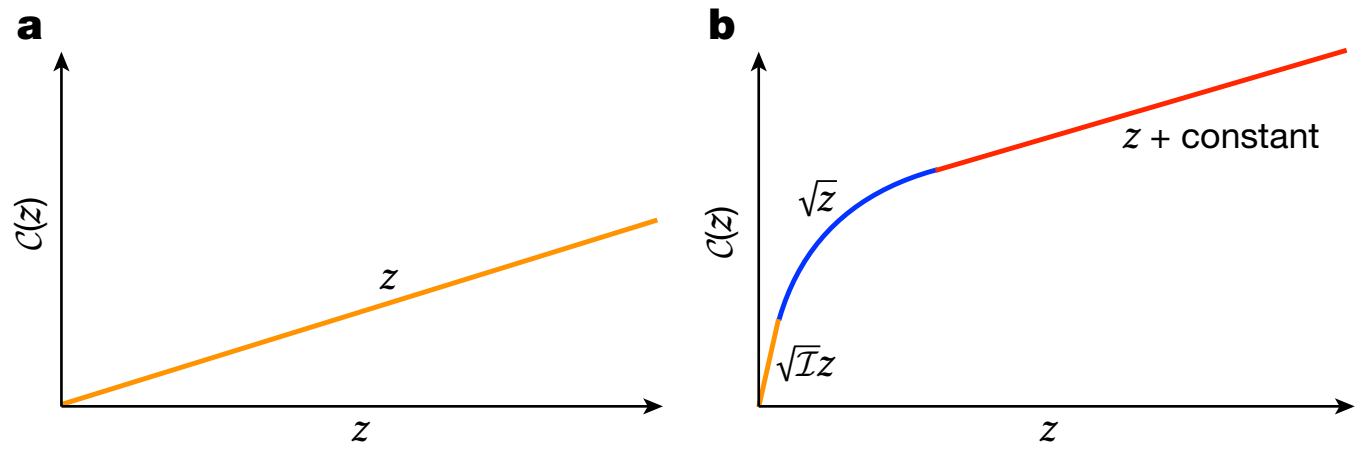

**Fig. 1 | Complexity of parallel parking a unicycle as a function of the distance *z* to the curb. a**, When $\mathcal{I}=1$, we drift directly into the parking spot, and the complexity is simply $z$. **b**, For large $\mathcal{I}$, there are three regimes: first, the complexity is linear with a large coefficient (orange); second, after the cut locus there is a square root regime (blue); and third, at large $z$, linear growth resumes with coefficient 1 (red).

Surprising fact: a small deformation can make the volume infinite. The surprise here is that in one sense, the homogeneous metric has been changed a lot (the volume becomes infinite), and in another sense, the metric has hardly been changed at all (the distance between any two points has a tiny additive variation). We will come to understand this phenomenon as an example of long-distance universality. Our small deformation will change the short-distance geometry, and we will see that while the volume is sensitive to the short-distance geometry, the distance function at large separations is not. The short-distance geometry and the long-distance geometry decouple.

## Low-dimensional Riemannian cases

### Berger sphere

Our first example of long-distance universality is the Berger sphere[9]. The Berger sphere manifests the surprising fact observed in the previous section and is the simplest non-trivial complexity geometry[3–7]—the complexity geometry of a single qubit[10,11]. It describes the difficulty of synthesizing elements of the group SU(2) when $\hat{\sigma}_x$ and $\hat{\sigma}_y$ rotations are cheap but $\hat{\sigma}_z$ rotations are expensive[11]. The metric in Euler angles $U=\mathrm{e}^{\mathrm{i}\sigma_z z}\mathrm{e}^{\mathrm{i}\sigma_y y}\mathrm{e}^{\mathrm{i}\sigma_x x}$ is

$$\mathrm{d}s^2=\cos^2 2y\,\mathrm{d}x^2+\mathrm{d}y^2+\mathcal{I}(\mathrm{d}z+\sin 2y\mathrm{d}x)^2. \quad (1)$$

For $\mathcal{I}=1$, this gives the standard round metric on a three-sphere—which is both homogeneous and isotropic—otherwise known as the bi-invariant inner-product ('Killing') metric on SU(2). For $\mathcal{I}\neq 1$, this gives a squashed three-sphere, which is still homogeneous but no longer isotropic (right invariant but no longer left invariant).

To exemplify the surprising fact, take the Berger sphere with $\mathcal{I}=10^{30}$, and make the small deformation $\mathcal{I}\to\infty$. This sends the volume to infinity, $\int\sqrt{\det[g]}=2\pi^2\sqrt{\mathcal{I}}\to\infty$, but makes only a tiny additive change to distances. It may seem surprising that taking $\mathcal{I}\to\infty$ only changes the distances a little, as it makes moving directly in the $\sigma_z$ direction infinitely expensive: the cost of directly synthesizing $\mathrm{e}^{\mathrm{i}\sigma_z z}$ is $\sqrt{\mathcal{I}}z$. But we can also synthesize $\mathrm{e}^{\mathrm{i}\sigma_z z}$ indirectly by circling in the $\sigma_x$ and $\sigma_y$ directions and using the group commutator:

$$(1+\mathrm{i}\sqrt{z}\sigma_x)(1+\mathrm{i}\sqrt{z}\sigma_y)(1-\mathrm{i}\sqrt{z}\sigma_x)(1-\mathrm{i}\sqrt{z}\sigma_y)\approx 1+\mathrm{i}^2 z[\sigma_x,\sigma_y]=1+2\mathrm{i}z\sigma_z. \quad (2)$$

The cost $\mathcal{C}$ of this indirect technique—the geometric length of this path—is independent of $\mathcal{I}$; because the amount of $\sigma_z$ generated is proportional to the area, the cost scales such as $\sqrt{z}$ at small $z$, and in general is[12,13]

$$\mathcal{C}_{\mathcal{I}=\infty}(\mathrm{e}^{\mathrm{i}\sigma_z z})\equiv\lim_{\mathcal{I}\to\infty}\mathcal{C}(\mathrm{e}^{\mathrm{i}\sigma_z z})=\sqrt{z(2\pi-z)}. \quad (3)$$

Let us see how much distances change under our small deformation. As there are two ways to manufacture $\sigma_z$, the line $\mathrm{e}^{\mathrm{i}\sigma_z z}$ is cut into two. In the inner region, direct synthesis is the cheapest so $\mathcal{C}(z)=\sqrt{\mathcal{I}}z$; in this region, the geometry depends strongly on the value of $\mathcal{I}$, and increasing $\mathcal{I}$ can make a large multiplicative change to distances. But when $\mathcal{I}$ is very big, this region is very small, extending only as far out as the cut locus at $\mathcal{C}_{\mathrm{cut}}\approx\mathcal{I}^{-1/2}$. For larger $z$, the optimal path is a mixture of direct and indirect synthesis, and the farther we go the less direct synthesis is involved. This means that the farther we go into the outer region, the less the distance function depends on $\mathcal{I}$: if you already were not availing yourself much of the direct technique, making the direct technique even more expensive makes little difference. This insensitivity to $\mathcal{I}$ reaches its apotheosis when making the very hardest unitary, $z=\pi$: for $\mathcal{I}\geq 1$, the cost of making $U=\mathrm{e}^{\mathrm{i}\sigma_z\pi}$ does not depend on $\mathcal{I}$ at all, because we can reach the antipode by proceeding along any great circle $\mathrm{e}^{\mathrm{i}\sigma_x\pi}=\mathrm{e}^{\mathrm{i}\sigma_y\pi}=\mathrm{e}^{\mathrm{i}\sigma_z\pi}=-\mathbb{1}$, so the greatest separation on the Berger sphere is exactly π and is completely insensitive to $\mathcal{I}$ (refs. 12,13). We are thus left with the following picture. Close to the origin, the distance function depends strongly on $\mathcal{I}$, but this region is small and shrinks to zero in the sub-Riemannian limit, $\mathcal{I}\to\infty$. Outside this region, distances are largely independent of $\mathcal{I}$. Nowhere does $\mathcal{I}$ make a large additive difference.

A careful analysis[12,13] confirms this picture and shows that the very largest additive discrepancy from the $\mathcal{I}=\infty$ distance is found near the cut locus, so for all $U$ and $\mathcal{I}\geq 1$,

$$\mathcal{C}_{\mathcal{I}=\infty}(U)-\mathcal{C}_{\mathcal{I}}(U)\leq O(\mathcal{I}^{-\frac{1}{2}}). \quad (4)$$

As $(10^{30})^{-\frac{1}{2}}=10^{-15}$, the $\mathcal{I}=\infty$ and $\mathcal{I}=10^{30}$ Berger spheres agree on distances to within a picometre. (A two-dimensional example that shows the same phenomenon is given in Supplementary Information 3).

Finally, let us examine the role of curvature. At large $\mathcal{I}$, the metric becomes strongly curved: the easy–easy section becomes very negatively curved $\kappa(x,y)=4-3\mathcal{I}$, and the easy–hard sections become very positively curved $\kappa(x,z)=\mathcal{I}$. We call the $\sigma_x$ and $\sigma_y$ directions easy because they are cheap to move it, whereas the $\sigma_z$ direction is hard to move in for $\mathcal{I}>1$. The curvature length $|\kappa|^{-1/2}\approx\mathcal{I}^{-1/2}$, which is also the distance to the cut locus in the hard direction, becomes very short. The high curvature explains how the metric can hide lots of volume at short distances that are invisible at long distances. Consider an operational definition of volume that counts how many marbles can be packed into the space: if we can cram in $n(r)$ marbles each of radius $r$, then the volume is proportional to $\lim_{r\to 0}r^3 n(r)$. Although the volume grows without bound as $\mathcal{I}\to\infty$, the effective volume $n(r)r^3$ at any finite value of $r$ does not. Instead, the effective volume grows like $r^{-1}$ as we take $r$ smaller and only levels off at $\sqrt{\mathcal{I}}$ once $r$ is less than the curvature length $\mathcal{I}^{-1/2}$. Thus even as $\mathcal{I}\to\infty$ the effective volume, as probed by experiments with finite resolution, stays finite.

### Euclidean group

Consider parallel parking a unicycle.

The unicycle starts facing parallel to the curb and ends facing parallel to the curb but displaced sideways by $z$. The configuration space of the unicycle is its possible locations $\{y,z\}$ and orientations $\{\theta\}$, forming the Euclidean group SE(2). There are three primitive operations: roll forwards or backwards; turn; or drift sideways (perpendicular to the rolling direction). We model the difficulty of any parking manoeuvre with

$$\begin{aligned}\mathrm{d}s^2&=\mathrm{d}x_{\parallel}^2+\mathcal{I}\mathrm{d}x_{\perp}^2+\mathrm{d}\theta^2\\&=(\cos\theta\mathrm{d}y+\sin\theta\mathrm{d}z)^2+\mathcal{I}(\sin\theta\mathrm{d}y-\cos\theta\mathrm{d}z)^2+\mathrm{d}\theta^2.\end{aligned} \quad (5)$$

The cost of parking $\mathcal{C}(z)$ is the length $\int\mathrm{d}s$ of the shortest path that connects our starting configuration $\{\theta,y,z\}=\{0,0,0\}$ to our parking spot $\{0,0,z\}$.

The important difference from the Berger sphere example is that, because there is no bound on how far away the curb can be, there is now no upper bound on $\mathcal{C}(z)$. However, we still get an inequality like equation (4) that is all the more powerful in this non-compact setting. Whether drifting is just as easy as rolling ($\mathcal{I} = 1$) or drifting is completely forbidden ($\mathcal{I} = \infty$), the cost of parking never changes by more than $O(1)$, so at large $z$ all metrics with $\mathcal{I} \geq 1$ have the same linear growth.

Figure 1 plots the cost of parking. For $\mathcal{I} = 1$, the metric is flat $\mathbb{R}^2 \times S^1$ and the optimal parking manoeuvre is just to drift directly in the $z$ direction at cost $\mathcal{C}(z) = z$. For large $\mathcal{I}$, the optimal parking manoeuvre can be more complicated, and there are three regimes. The first regime, at tiny $z$, is to drift directly in the $z$ direction, which gives linear growth $\mathcal{C}(z) = \sqrt{\mathcal{I}}z$ that is strongly dependent on $\mathcal{I}$; this regime extends as far as the cut locus, a curvature length from the origin at $\mathcal{C}_{\text{cut}} \approx \mathcal{I}^{-1/2}$. The second regime involves commuting the two easy directions giving $\mathcal{C}(z) \approx \sqrt{z}$. (Turn through an angle $\sqrt{z}$, roll forwards $\sqrt{z}$, turn back through $-\sqrt{z}$ to be again parallel to the curb and then roll backwards $-\sqrt{z}$ into the parking spot). So far this is the same as the Berger sphere example, but the difference is what happens next. For $z \gtrsim 1$, the square root behaviour comes to an end, because once you have turned to face the $z$ direction, there are no further efficiencies from turning more. But because we have already paid the fixed cost of turning to face the $z$ direction, the marginal cost of going farther in the $z$ direction is just the cost of rolling forwards. The third regime is thus another linear regime, but this time with gradient one, $\mathcal{C}(z) = z + \text{constant}$.

There is a simple upper bound on the distance that works for all values of $\mathcal{I}$. Consider a strategy in which we first turn 90°, then roll forwards all the way to the curb, and then on arrival turn 90° to end parallel to the curb. This upper-bounds $\mathcal{C}(z) \leq \frac{\pi}{2} + z + \frac{\pi}{2}$, and by optimizing this strategy we find that at large $z$

$$\mathcal{C}(z) = z + 2\sqrt{1 - \mathcal{I}^{-1}} + \ldots, \tag{6}$$

in which the omitted terms vanish rapidly at large $z$. Indeed, a more careful analysis finds the behaviour shown in Fig. 1.

We have thus found a large universality class of metrics (every $\mathcal{I} \geq 1$), all of which agree at long distances, up to moderate additive corrections. In this equivalence class, the leading-order long-distance behaviour is independent of $\mathcal{I}$. As the dependence on $\mathcal{I}$ appears to be only in the sub-leading corrections, we may say that $\mathcal{I}$ is an irrelevant deformation. Although the members of the equivalence class agree at long distances (borrowing the language of quantum field theory, we could say they agree in the IR), they disagree markedly at short distances (in the UV): they have radically different curvatures, and radically different cut loci.

For most members of the universality class, the relationship between short- and long-distance geometry is convoluted. For one member, however, the relationship is straightforward. For $\mathcal{I} = 1$—the critical value below which changes in $\mathcal{I}$ do affect the long-distance behaviour—the UV behaviour and the IR behaviour match exactly, with the same linear growth coefficient. We refer to this special value as giving the critical metric. If you wish to approximate distances in the IR, it is easiest just to calculate with this critical metric, no matter what the true UV value of $\mathcal{I} \geq 1$ is. As we approach the critical metric, the cut locus gets pushed out to $z = \infty$ and the curvature becomes small. We use these properties to help identify the critical metric in more complicated models.

Finally, let us comment on the error, or lack thereof. For the unicycle parking, we did not have to introduce a tolerance, because we can hit exactly with minimal extra cost any point that we can get close to, even in the $\mathcal{I} \to \infty$ limit. From the IR point of view, this is obvious—because small changes in direction have small costs, and we have plenty of 'wiggle room' to make minor adjustments to the end point. From the UV point of view, this is surprising—a Suzuki–Trotter-style[14] perturbative expansion in $z$ that models the path as a piecewise linear sequence of time-independent Hamiltonians markedly overestimates the cost of correcting errors.

The Berger sphere has a diameter π and therefore no long-distance regime. The unicycle parking example does have a long-distance regime, but the critical geometry is flat. In the next example, we will see a non-trivial critical distance function.

## High-dimensional gate example

In the gate model of quantum complexity, we build complex unitaries by arranging simple unitaries in a circuit[15]. The simple unitaries are $k$-local gates, which means unitaries in $U(2^k)$ that act non-trivially on $k$ qubits. We can use any element of $U(2^k)$, at a cost $\sqrt{\mathcal{I}_k}$ (the cost therefore depends on only $k$); the cost of a circuit that uses $n_k$ $k$-local gates is $\mathcal{C}[\text{circuit}] = \sum_k \sqrt{\mathcal{I}_k} n_k$. To complete our definition of complexity, we need to specify the penalty schedule, which means specifying the penalty factors $\mathcal{I}_k$ for every value of $k \leq N$. We then ask how the complexity of a given unitary depends on our choice of penalty schedule.

Our first observation is that if we take any one of the $\mathcal{I}_k$ large (while keeping the others fixed) the complexity soon becomes completely independent of the value of that penalty factor. This is because instead of directly deploying a $k$-local gate, we could also indirectly synthesize the same unitary with a subcircuit built out of cheaper gates. No gate is indispensable, and the set of $m$-local gates for any individual $m \geq 2$ is sufficient to reconstitute all the others[15]. To replace a $k$-local gate we never need to pay more than $\min_m \sqrt{\mathcal{I}_m} n_m(k)$, where $n_m(k)$ is defined as the number of $m$-local gates needed to build any $U \in U(2^k)$. Once $\sqrt{\mathcal{I}_k}$ exceeds this critical value, the complexity becomes completely independent of $\sqrt{\mathcal{I}_k}$.

Now let us consider the critical schedule for which every penalty factor takes its critical value. This means that the price of each gate is the same as the cost of indirectly synthesizing it out of cheaper gates. We can show that there is a schedule with approximately this property by estimating $n_m(k)$, which simple dimension counting bounds by

$$n_m(k) \equiv \text{ number of } m\text{-locals to build } U(2^k) \geq \frac{\dim[U(2^k)]}{\dim[U(2^m)]} = 4^{k-m}. \tag{7}$$

This lower bound is approximately saturated[16] (although not exactly saturated because when two consecutive gates overlap on a qubit, there is an U(2) isotropy subgroup, so adding the dimensions of the two gates overcounts by $\dim[U(2)] = 4$). If we fix the normalization by setting $\mathcal{I}_2 = 1$, this means that the critical schedule is roughly

$$\overline{\mathcal{I}}_k \approx 4^{2(k-2)}, \tag{8}$$

independent of $N$ for fixed $k$. For this critical schedule, direct and indirect syntheses are approximately degenerate. If we start with the critical schedule and make one of the penalty factors more expensive, this has little effect on the complexity of any unitary, because we will just switch to the cheaper option. By contrast, if we start with the critical schedule and make any penalty factor less expensive, the complexity of almost all unitaries reduces.

We have thus, once again, found a large universality class of definitions of complexity (every schedule with $\mathcal{I}_2 = \overline{\mathcal{I}}_2 = 1$ that is no less expensive than the critical schedule $\forall k, \mathcal{I}_k \geq \overline{\mathcal{I}}_k$), all of which approximately agree on the complexities of all unitaries. If the penalty schedule is in this universality class, the large-distance complexity will be the same. For most members of this class, the cheapest way to make a typical element of $U(2^k)$ involves a convoluted compilation strategy, but there is a unique member of the class for which the optimal compilation is straightforward: for the critical schedule, we make the element with a single gate.

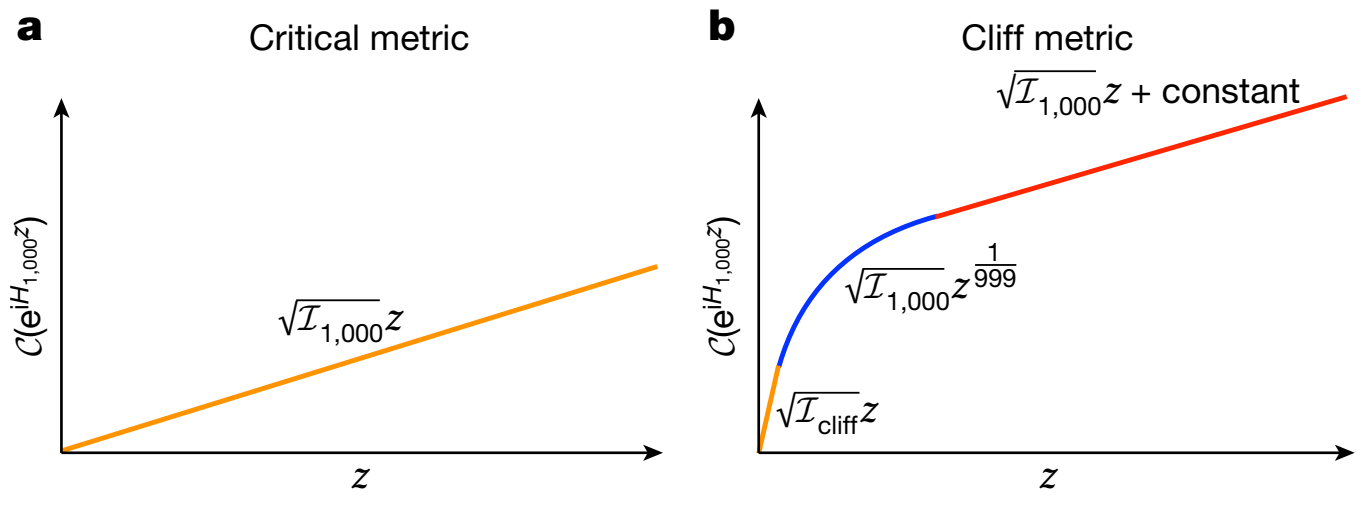

**Fig. 2 | Conjectured complexity $\mathcal{C}$ as a function of the inner-product distance $z$, in which $H_{1,000}$ is a 1,000-local polynomial Hamiltonian. a**, For the proposed critical metric, the complexity grows linearly with the same coefficient at all $z$ (until it saturates at a $z$ exponential in $N$). **b**, For the cliff metric, equation (9), at very short distances the complexity grows linearly (orange), then it hits the cut locus and slows to sublinear growth (blue), before transitioning to linear growth again but with a lower slope that matches the critical metric (red).

The universality behaviour is somewhat broader than the class $\mathcal{I}_2 = \bar{\mathcal{I}}_2$, $\mathcal{I}_k \geq \bar{\mathcal{I}}_k$. First, if we start with the critical schedule and make $\mathcal{I}_2$ bigger, this may have a large effect at short distances but only a small effect at long distances: in the language of quantum field theory, this would be an 'irrelevant' deformation. Similarly, if we take some of the $\mathcal{I}_k$ with moderate $k$ and make them cheaper than the critical schedule, this would affect both the UV and the IR, but its effect on the IR would be just to multiply all complexities by $\min_k \sqrt{\mathcal{I}_k/\bar{\mathcal{I}}_k}$: the entire effect on the IR of a complicated UV deformation would be summarized by a single parameter.

The features mentioned above all have direct analogues in the examples given in the section 'Low-dimensional Riemannian cases'. Now let us examine a feature that emerges only when the number of dimensions is large. The critical schedule was defined by demanding that $\sqrt{\bar{\mathcal{I}}_k} = \min_{m<k} \sqrt{\bar{\mathcal{I}}_m}\, n_m(k)$. But there is a scaling symmetry $n_m(k) \approx n_m(p)\, n_p(k)$, which means that if we want to make a $k$-local unitary out of $m$-local gates, we pay little extra cost for doing a hierarchical compilation that first makes the $k$-local unitary out of $p$-local gates and then makes the $p$-local gates out of $m$-local gates. This scaling symmetry means that for the critical schedule the quantity $\sqrt{\bar{\mathcal{I}}_m}\, n_m(k)$ is largely independent of $m$. There is thus a huge degeneracy of ways to make a $k$-local gate: there is an equal-length path for every value of $m$, as well as a great many more paths of mixed $m$. It is this massive degeneracy and redundancy—called 'load balancing' in network engineering—that makes the numerical value of the complexity so robust against upward deformations of the penalty schedule.

## Application to complexity geometry

We now extrapolate the lessons learnt from the simple examples seen in the sections 'Low-dimensional Riemannian cases' and 'High-dimensional gate example' and to make conjectures about the long-distance behaviour of high-dimensional complexity geometry. Complexity geometry is reviewed in the Methods.

### Main conjectures

Let us ask how the value of the complexity of a unitary depends on the choice of penalty schedule. Our fundamental observation is that the complexity geometry shares the same properties that drove the universal behaviour we saw in the previous examples. The universal behaviour in these cases was caused by the overcompleteness of the set of primitive operations, which meant that there were many different ways to effect any given change. The complexity geometry is even more overcomplete than the gate definition of the previous section, because on the one hand its target is the same (elements of $U(2^N)$) but on the other hand the tools at its disposal are much more powerful: whereas a $k$-local gate is constrained to act only on $k$ qubits at a time, in the complexity geometry we may also move in any polynomial superposition over different $k$-local terms, or terms of mixed $k$-locality; and whereas the gate definition gets charged a full $\sqrt{\mathcal{I}_k}$ for even a small step of inner-product size $\epsilon$ in a $k$-local direction, the complexity geometry charges only $\sqrt{\mathcal{I}_k}\,\epsilon$ and is, therefore, able to change direction without penalty and economically deploy very wiggly paths. Because constructing a path through the complexity geometry affords so many more options than compiling gates into a circuit, the primitive operations are more overcomplete, and so the universality behaviour should be correspondingly more robust.

On the basis of these considerations, we expect an enhanced version of the same universality properties we saw earlier. We can formalize this into two (independent) conjectures:

**Conjecture 1.** There exists a critical schedule $\bar{\mathcal{I}}_k$, and a universality class consisting of all schedules that are anchored at $\mathcal{I}_2 = \bar{\mathcal{I}}_2$ that are no easier than the critical schedule, $\forall k$, $\mathcal{I}_k \geq \bar{\mathcal{I}}_k$. In this universality class, the distance functions may differ greatly at short separation (in the UV), but will approximately agree at long separation (in the IR).

**Conjecture 2.** The critical schedule $\bar{\mathcal{I}}_k$ is the unique member of the universality class for which the UV and the IR behaviours match. This means that for the critical schedule the cut locus is pushed far out in almost all directions, so that geodesics leaving the origin typically remain minimal for a time exponential in $N$ and the linear growth continues uninterrupted with the same coefficient at long and short distances.

To make these conjectures more quantitative, we need to specify how close the members of the universality class are to each other and to identify the critical schedule. Our purpose in this paper is not to settle these more quantitative questions but to lay out an effective geometry program to address them, and to identify some valid candidate answers. The easiest schedule in the universality class is the critical schedule, $\bar{\mathcal{I}}_k$, whereas the hardest schedule is the cliff metric:

$$\mathcal{I}_1 = \mathcal{I}_2 = 1 \quad \text{and} \quad \mathcal{I}_{k\geq 3} = \mathcal{I}_{\text{cliff}} \tag{9}$$

in the limit $\mathcal{I}_{\text{cliff}} \to \infty$. As the penalty factors vary by an infinite amount between these two extremes, we might think that the assigned complexities could as well; the content of conjecture 1 is that this does not happen. Let us now describe a plausible quantitative conjecture inspired by the results of the previous sections and describe some supporting evidence. In the Methods, we lay out some more conjectures about the critical metric and describe more directly the relationship to the existing mathematics literature.

### Quantitative conjecture for complexity growth

For concreteness, let us consider the complexity of a unitary of the form $U = e^{iH_k z}$, where $H_k$ is a typical (polynomial) $k$-local Hamiltonian. We normalize such that $\mathrm{Tr}H^2 = 1$, so that $z$ is the inner-product (Killing) distance. This is a useful case to consider because constant $H_k$ gives a geodesic. This is because all terms in $H_k$ have the same weight, and by assumption the penalty factor depends only on the weight, and a constant Hamiltonian in which all terms have the same penalty factor gives a geodesic. A candidate quantitative conjecture is shown in Fig. 2.

The complexity growth for the easiest schedule in the universality class, the critical schedule $\bar{\mathcal{I}}_k$, is shown in Fig. 2a. For a $k$-local Hamiltonian, our statement that the critical metric typically does not hit a cut locus means the complexity $\mathcal{C}$ is simply given by the direct geodesic distance $s$, that is, $\mathcal{C}(e^{iH_k z}) = s = \sqrt{\bar{\mathcal{I}}_k}\, z$, until $z$ is exponentially large in $N$.

The complexity growth for the hardest schedule in the universality class, the infinite-cliff metric $\mathcal{I}_1 = \mathcal{I}_2 = 1$, $\mathcal{I}_{k\geq 3} = \mathcal{I}_{\text{cliff}} \to \infty$, is shown in Fig. 2b. At vanishingly small distances, the two schedules disagree about the complexity of $e^{iH_k z}$ by a multiplicative factor of $\sqrt{\mathcal{I}_{\text{cliff}}/\bar{\mathcal{I}}_k} \to \infty$. However, as $\mathcal{I}_{\text{cliff}}$ diverges, the cut locus in the $z$ direction approaches the origin, and beyond the cut locus $e^{iH_k z}$ will be more economically synthesized by an indirect path that commutes the cheap directions. For the cliff metric, we expect the same three-regime behaviour as we saw for the parking unicycle in Fig. 1. The first regime is linear with a huge coefficient $\mathcal{C} = \sqrt{\mathcal{I}_{\text{cliff}}}\,z$. This regime soon ends at a cut locus (and is squeezed out entirely for $\mathcal{I}_{\text{cliff}} \to \infty$). Beyond the cut locus, we synthesize $e^{iH_k z}$ indirectly, by commuting 2-local Hamiltonians. As the number of 2-local operators we must commute to make a $k$-local operator is at least $k - 1$, the second regime has sublinear growth with $\mathcal{C} \propto z^{\frac{1}{k-1}}$ with some multiplicative coefficient (the exponent of this power law is fixed by the ball-box theorem)[17]. After $z \sim 1$, linear growth resumes with coefficient given by the critical metric. In a strong form of the conjecture, the additive deviation of the complexity is never more than the cost of a single $k$-local gate so that $\mathcal{C}_{\text{cliff}}\,[e^{iH_k z}] \le \mathcal{C}_{\text{critical}}\,[e^{iH_k z}] + O(\sqrt{\mathcal{I}_k})$.

We have argued that there exists a critical metric. In the section 'Quantitative conjecture for critical metric', we make a quantitative conjecture for the form of this metric, and argue that the penalty factors should grow at least exponentially with the number of qubits they touch. In the section 'Consistency checks', we discuss further evidence for these conjectures.

## Discussion

In this paper, we investigated long-distance universality and argued that short-distance geometry and long-distance geometry can decouple. For homogeneous metrics, the short-distance properties—those that can be measured in the vicinity of a point, such as the curvature and its derivatives—determine (up to global identifications) everything about the entire geometry. But the farther away from the origin you get, the more convoluted the relationship between short-distance and long-distance geometry becomes. We illustrated this by considering firing a geodesic in some direction and following it for a distance $z$. At first, the distance from the origin is just $z$, so the distance function is simple. But when a cut locus is encountered, the geodesic is no longer minimal, and the distance function becomes much more complicated. Nevertheless, we argued that at even greater separations the distance function becomes simple again. We argued that far out beyond the cut locus a new kind of order emerges. This new order—an emergent long-distance metric—is largely insensitive to the details of the short-distance metric. There is instead a form of universality, in which a broad class of short-distance metrics all give rise to the same effective long-distance geometry, with the entire effect of the short-distance geometry being summarized in a handful of relevant parameters. This is the short–long decoupling that enables different spaces of wildly different volumes to agree on every distance to within a picometre.

In the section 'Low-dimensional Riemannian cases', we exhibited long-distance universality in low-dimensional Riemannian examples that we could solve exactly. In the section 'High-dimensional gate example', we exhibited large-complexity universality in the high-dimensional but non-Riemannian case of gate complexity. In the section 'Application to complexity geometry' we studied high-dimensional Riemannian geometries, and marshalled the evidence for our conjectures that these complexity geometries exhibit long-distance universality (which in this context is, by definition, the same thing as high-complexity universality).

Our investigations were originally motivated by the ideas in holographic black holes. Although our results stand independent of those motivations, we discuss the implications for holography in the section 'Relation to black holes and holography'; we observe that if our conjectures are true, then complexity geometry may provide just the precise-but-robust definition of complexity needed to undergird the holographic complexity correspondences[18–21]. An inspiration for our investigations was the Wilsonian theory of renormalization in quantum field theory[1]. In some sense, this paper is an attempt to apply those ideas to geometry. We develop the connection further in the section 'Wilsonian connections'. Although we have emphasized the application of our ideas to complexity geometry, it seems likely that they are applicable to all right-invariant metrics on sufficiently 'large' Lie groups, where 'large' could mean either non-compact or a sequence of compact Lie groups with growing dimensions; an interesting mathematical program (Methods and Supplementary Information) would be to characterize the equivalence classes of right-invariant geometries on large Lie groups.

Long-distance universality should be a robust feature of sufficiently rich high-dimensional spaces. Here we have examined the implications for one example of a high-dimensional space: the space of unitary functions on quantum states, which has a dimension exponential in the number of qubits. But high-dimensional spaces are found in many areas in physics and computer science, from many-body systems to deep neural networks. It is tempting to speculate that the concept of emergent geometry may be of broader relevance.

## Online content

Any methods, additional references, Nature Portfolio reporting summaries, source data, extended data, supplementary information, acknowledgements, peer review information; details of author contributions and competing interests; and statements of data and code availability are available at https://doi.org/10.1038/s41586-023-06460-3.

# Methods

## Review of complexity geometry

Now we provide an overview of Nielsen's complexity geometry[3–7]; we recommend ref. 11 (sections 1 and 2) as a more pedagogical review (for other recent works, see refs. 22–29). Similar to gate complexity, complexity geometry endows the unitary group $U(2^N)$ with a right-invariant distance function $\mathcal{C}(U_1, U_2) = \mathcal{C}(U_1U_2^{-1}, \mathbb{1})$ between two unitaries, which we interpret as the definition of the relative complexity. However, in contrast with gate complexity, which is not a continuous function of $U$, complexity geometry endows the unitary group with a smooth Riemannian metric. A general Riemannian right-invariant metric is parameterized by a symmetric moment of inertia tensor $\mathcal{I}_{IJ}$, so that the infinitesimal distance $\mathrm{d}s$ between $U$ and $U + \mathrm{d}U = (\mathbb{1} + \mathrm{i}\sigma_I\mathrm{d}\Omega^I)U$ is given by

$$\mathrm{d}s^2 = \mathrm{d}\Omega^I \mathcal{I}_{IJ} \mathrm{d}\Omega^J \tag{10}$$

and $\mathrm{d}\Omega^I = \mathrm{iTr}\ \mathrm{d}U^\dagger \sigma_I U$. Here $\sigma_I$ are the generalized-Pauli operators, which provide a complete basis on the tangent space,

$$\{\sigma_I\} \in \{\mathbb{1}, \sigma_a^{(A)}, i\sigma_a^{(A)}\sigma_b^{(B)}, \sigma_a^{(A)}\sigma_b^{(B)}\sigma_c^{(C)}, \ldots\}, \tag{11}$$

where lowercase letters run over the Pauli indices $a \in \{x, y, z\}$ and capital letters indicate the qubit on which the Pauli operator acts, $A \in \{1, 2, \ldots, N\}$, and we have normalized the trace so that $\mathrm{Tr}\mathbb{1} = \mathrm{Tr}\sigma_I^2 = 1$. The distance between the two unitaries is defined as the minimal geodesic distance in this metric, $\mathcal{C} = \min \int \mathrm{d}s$. If $\mathcal{I}_{IJ}$ were the identity matrix, this would recover the standard inner-product metric on the unitary group in which all directions are equally easy to move in, but in general a complexity geometry will stretch difficult directions to make complex unitaries farther away. To specify a complexity geometry, we must specify $\mathcal{I}_{IJ}$. Following Nielsen, let us consider $\mathcal{I}_{IJ}$ s that are diagonal in the generalized-Pauli basis and for which the penalty factor of a given generalized-Pauli operator is solely a function of the $k$-locality, that is, solely a function of the weight (or size) $k$ of the operator, defined as the number of capital indices in equation (11). Notice that a $k$-local Hamiltonian can be an arbitrary superposition of weight-$k$ generalized-Pauli operators—it is allowed to touch all the qubits, so long as no single term touches more than $k$ at a time—whereas a $k$-local gate (defined in the section 'High-dimensional gate example') acts on only $k$ qubits. To specify the metric, we then need only to specify the penalty schedule $\mathcal{I}_k$, that is, the choice of $\mathcal{I}_k$ for each $k \le N$.

We can also consider the complexity geometry on $2N$ Majorana fermions; this would be natural for studying the complexity of the Sachdev–Ye–Kitaev (SYK) model[30,31] or other fermionic theories[32,33].

## The critical metric

Here we elaborate on the conjectures made in the section 'Application to complexity geometry'.

**Quantitative conjecture for critical metric.** Let us identify a good candidate to be the critical schedule. Unlike the cliff metric, for which all $\mathcal{I}_{k\ge3}$ are the same, for the critical metric we expect the penalty factors to steadily grow with $k$. This is to reflect the fact that the difficulty of compiling a direction using two-local Hamiltonians increases with the $k$-locality. On the other hand, we expect the largest penalty factor to be exponentially large in $N$. This is to reflect the fact that the maximum complexity is exponentially large in $N$, and the maximum complexity is bounded above by the maximum penalty factor, $\text{complexity}_{\max} \le \pi\sqrt{\mathcal{I}_{\max}}$. In equation (8), we showed that the exponential metric,

$$\overline{\mathcal{I}}_k \approx 4^{2(k-2)}, \tag{12}$$

is a good approximation to the critical schedule of the gate model, up to sub-exponential corrections. Our quantitative conjecture is that an exponential metric, possibly with some different base $x$ not necessarily equal to 4, is also a good approximation to the critical schedule for the complexity geometry,

$$\overline{\mathcal{I}}_k \approx x^{2(k-2)}. \tag{13}$$

A previous study[7] pointed out some of the attractive features of the exponential metric for complexity geometry. One of the features is that, similar to the critical metric we examined in the section 'Low-dimensional Riemannian cases', the exponential metric has low curvature. Let us review that now. Another study[2] showed that when the commutator of two directions is much more expensive than either direction individually, the sectional curvatures are

$$\kappa(H_I, H_J) \approx -\frac{\mathcal{I}_{[H_I,H_J]}}{\mathcal{I}_{H_I}\mathcal{I}_{H_J}},\ \kappa(H_I, [H_I, H_J]) \approx +\frac{\mathcal{I}_{[H_I,H_J]}}{\mathcal{I}_{H_I}\mathcal{I}_{H_J}}. \tag{14}$$

Two generalized-Pauli operators have a non-zero commutator only when they overlap on at least a single qubit, so the weight of the commutator is always less than the sum of the weights of the two operators,

$$\text{Weight}([\sigma_I, \sigma_J]) \le \text{Weight}(\sigma_I) + \text{Weight}(\sigma_J) - 1. \tag{15}$$

This means that for the exponential metric the magnitude of all the sectional curvatures, of both signs, is always less than $O(1)$. In the section 'Low-dimensional Riemannian cases', we saw that low curvature is a signature of the critical metric. By contrast, the cliff metric has huge sectional curvatures because two easy 2-local directions ($\mathcal{I}_2 = 1$) commute to a very hard 3-local direction ($\mathcal{I}_3 = \mathcal{I}_{\text{cliff}}$). This huge sectional curvature of the cliff metric indicates that the cut locus in the hard direction is close.

**Consistency checks.** Now let us describe some important consistency checks on these ideas.

An important consistency check is the diameter. If members of the universality class are to have approximately the same long-distance behaviour, then they certainly need to approximately agree on the diameter (that is, the greatest separation of any pair of points). We saw in the Berger sphere example that all members of that universality class agree on the diameter exactly. It is not obvious in advance that the cliff metric with $\mathcal{I}_{\text{cliff}} \to \infty$ should even have a finite diameter, because some of the directions are becoming infinitely expensive and the volume is diverging. However, Chow's theorem[34,35] ensures that so long as we can reach every element of the algebra by nested commutators of finitely expensive elements of the algebra, then the distance function converges in the limit $\mathcal{I}_{\text{cliff}} \to \infty$ and the diameter is therefore finite. We can place a tighter upper bound by noticing that everything we can do in the gate definition of complexity from the section 'High-dimensional gate example' we can do no more expensively (up to a multiplicative factor of π) in the complexity geometry with the same penalty schedule because every $k$-qubit gate $U(2^k)$ can be made by evolving with a $k'$-local Hamiltonian ($k' \le k$) that acts only on those $k'$ qubits for an inner-product distance at most π, giving a complexity geometry cost at most $\pi\sqrt{\mathcal{I}_k}$. Furthermore, we know from ref. 16 that even with the infinite-cliff schedule we can construct a circuit for every element of $U(2^N)$ with a cost no greater than $N^2 4^N$. This gives the upper bound. A previous study[4] was also able to prove a lower bound on the diameter of the cliff metric of $4^{N/3}$. If our conjecture is correct, the diameter of the critical schedule cannot be substantially less than the diameter of the infinite-cliff metric. It is therefore relevant that in ref. 36 a result is proved that lower-bounds the diameter of the exponential metric, equation (13), for all $x > 1$ (and several other metrics) by a quantity exponentially large in $N$. Our conjecture thus passes this consistency check.

This result is encouraging, but much weaker than what we want to show. We want to show that not only do all metrics in the universality class agree on the diameter, but also they approximately agree on the complexity of almost all sufficiently complex unitaries. Let us now report a step in that direction.

First, let us describe a heuristic compilation strategy for $e^{iH_k z}$ that suggests an upper bound for the critical schedule. This compilation strategy aims to synthesize $e^{iH_k z}$ using only 2-local Hamiltonians (which are always cheap for all members of the universality class). A typical $k$-local Hamiltonian $H_k = \sum_I \omega_I \sigma_I$ is a weighted sum of about $3^k \binom{N}{k}$ $k$-local generalized-Pauli operators (monomials). The dimensionality of the space of $k$-local Hamiltonians is therefore exponentially bigger than the dimensionality of the space of 2-local Hamiltonians, by a factor of

$$n_2(k) \equiv \frac{3^k \binom{N}{k}}{3^2 \binom{N}{2}}. \tag{16}$$

If we wish to write a typical $k$-local Hamiltonian as the nested commutator of 2-local Hamiltonians, simple dimension counting tells us that this requires no fewer than $n_2(k)$ levels of nesting. However, there are atypical $k$-local Hamiltonians that can be generated much more compactly. In particular, there is a special set of Hamiltonians, of dimension approximately $(k-1)3^2 \binom{N}{2}$, that can be written as the nested commutator of only $(k-1)$ 2-local terms. This set includes the $k$-local generalized-Pauli operators. Our compilation strategy uses these special Hamiltonians as building blocks. In particular, we use the fact that any operator of the form $e^{i\sigma_I z}$, where $\sigma_I$ is a $k$-local generalized-Pauli operator, can be constructed exactly out of 2-local operations with a cost no greater than $O(k)$.

An example of a compilation strategy is the following. Any generalized-Pauli $\sigma_K$ of weight $k$ can be written as the commutator of a weight-$(k-1)$ generalized-Pauli $\sigma_J$ and a weight-2 $\sigma_I$ that overlap at a single qubit. These three operators satisfy $e^{i\sigma_K z} = e^{i\sigma_I \frac{\pi}{4}} e^{i\sigma_J z} e^{-i\sigma_I \frac{\pi}{4}}$, just as they would if they were elements of SU(2). In this way, we can recursively synthesize motion in any $k$-local monomial direction with a cost $\mathcal{C}[e^{i\sigma_K z}] \le O(k)$. As moving indirectly in monomial directions is so cheap, the cut locus in monomial directions is very close to the origin even for the critical schedule. The extreme closeness of cut loci in monomial directions does not violate conjecture 2 because monomial directions are extremely atypical.

This implies that we can approximate the operator $U = \prod_I e^{i\omega_I \sigma_I z}$ with a total cost of about $\mathcal{C} \approx kn_2(k)$. This operator agrees with our target operator $e^{i\sum_I \omega_I \sigma_I z}$ at leading order in $z$, and has an inner-product error of about $z^2$. This can be improved to $z^3$ by using the next order in the Suzuki–Trotter expansion, but going to even higher orders becomes prohibitively expensive. It is at this point that we make our heuristic step. In the Euclidean group example, we saw that the complexity geometry has so many degrees of freedom that by making minor deformations of the path we can correct small errors at small extra cost, in a way that is not captured by any finite order of the Suzuki–Trotter expansion, and is instead an emergent feature in the IR. Compared with the SU(2) example in the section 'Berger sphere', the task of compiling in U($2^N$) is complicated by the fact that there are many more directions in which to err; on the other hand, there are correspondingly more directions in which we can wiggle the path to eliminate the error, and as a statistical matter, we expect that to dominate. If the small inner-product errors can be corrected by wiggling the path, then we can synthesize $e^{iH_k z}$ for $z < 1$ at cost $kn_2(k)$. To generate $e^{iH_k z}$ at larger values of $z$, the triangle inequality ($\mathcal{C}(az) \le a\,\mathcal{C}(z)$ for any $a \in \mathbb{N}$) guarantees that the complexity grows no faster than linearly with coefficient $kn_2(k)$. This argument heuristically shows that the binomial metric is in the same universality class as the infinite-cliff metric, and therefore upper-bounds the critical schedule:

$$\overline{\mathcal{I}}_k \lesssim k^2 n_2(k)^2. \tag{17}$$

The upper-bound equation (17) holds at all but the largest $k$, where the analysis becomes unreliable. Note also that although the binomial metric does not have a curvature as small as the exponential metric, it is still very moderate $|\kappa| \le O(N)$ compared to the cliff metric $|\kappa| \sim \mathcal{I}_{\text{cliff}}$. The reasoning that leads to equation (17) is heuristic, because to eliminate error it appeals to a statistical argument. In ref. 37, it is shown that there is a weaker result that can be proved. The study also shows that any unitary that can be reached with a path that in the binomial metric has a length $\mathcal{C}_{\text{bin.}}(U)$ can be approximated to within inner-product error $\epsilon$ by a path that in the infinite-cliff metric has a length

$$\mathcal{C}_{\text{cliff}}(U) \le 17\pi^2 N \frac{\mathcal{C}_{\text{bin.}}(U)^{7/2}}{\epsilon^{5/2}}. \tag{18}$$

Our conjectures imply that this can be improved from polynomial to linear-with-additive-constant and from approximate to exact.

Finally, let us note that a property we have conjectured for the complexity geometry—namely, linear growth of complexity that lasts for an exponential duration—has been proved already in two simple toy models: a discrete random-circuit model on Cayley graphs[38] and a continuous random-circuit model on the unitary group that tolerates zero error[39].

**Next steps.** In attempting to prove, refute or provide further evidence for our conjectures about precise equivalences between high-dimensional complexity geometries, two broad strategies could be pursued: starting at low dimension and working up or starting at high dimension and making the equivalencies more precise.

Following the latter strategy, we could initiate a program of proving increasingly precise equivalence relations. We would show that all metrics in the equivalence classes have approximately the same large-separation distance functions, with escalating strength for the form of the discrepancy (for example, polynomial versus linear versus additive), for the $N$ dependence of the discrepancy (for example, exponential versus polynomial versus linear), for the form of the error tolerated (for example, inner-product distance versus operator-norm distance versus exact), and for whether tight bounds on the discrepancy are to be found in only moderately easy directions or in all directions. This program would pursue a progressive strengthening of the results given in ref. 37.

A complementary program would be to start with the low-dimensional examples in the section 'Low-dimensional Riemannian cases' and steadily increase the dimension. For example, a concrete next step to test our conjectures would be to numerically calculate the distance function for a modest number of qubits (or Majoranas), extending the numerical analysis of ref. 40 from two qubits to a handful or more.

## Relation to black holes and holography

In the context of the gauge–gravity duality[41], it has been conjectured that some geometric properties of the black hole interior[18–21] are related to the quantum complexity of the holographic dual of the black hole. For example, in ref. 19 it was conjectured that

$$\text{Complexity} \approx \text{Volume}, \tag{19}$$

where the volume is the volume of a wormhole behind a black hole horizon and the $\approx$ symbol accounts for an unknown multiplicative constant. In refs. 20,21, an even more precise conjecture was made:

$$\text{Complexity} = \frac{\text{Action}}{\pi\hbar}, \qquad (20)$$

where again the action is evaluated for a certain geometric region of the holographic wormhole and this time there is no multiplicative ambiguity.

From the point of view of conventional complexity theory, equations (19) and (20) are alarming. On the right-hand side of the equations we have geometric quantities whose values can be calculated exactly, whereas on the left-hand side we have a quantity that in the conventional view is robustly defined only up to polynomial equivalence, and only then not for a single solution but for a family of solutions of varying $N$ in the limit that $N$ gets large. In this view, it is a category error to expect to be able to give robust meaning to the numerical value of the complexity of a particular unitary. Of course, even in this view, we can always extract a numerical value by being extremely precise about which choices we make for the definition of complexity (for example, exactly which primitive gates or which penalty factors), but there would be no expectation that the numerical value would be robust against perturbing these choices. Furthermore, there are no known principles that would dictate these seemingly arbitrary choices.

But if the conjectures in this study are correct, equations (19) and (20) are no longer so alarming. Instead, the universality of long-distance complexity tells you that (in the semi-classical limit, in which complexities are large and the dual spacetime is effectively classical), there is a robust definition of complexity to place on the left-hand sides of equations (19) and (20), in which many of the seemingly arbitrary choices of penalty factors do not matter. This could enable a rigorous formulation of holographic complexity.

Further links between holography and the results are discussed in Supplementary Information 2.

### Wilsonian connections

We make explicit the analogy between our findings in geometry and the Wilsonian theory of renormalization[1].

A starting point for complexity geometry, both logically and historically, is Nielsen's cliff metric with a huge penalty factor for the non-easy directions $\mathcal{I}_{\text{cliff}}$ (see section 'Main conjectures'). In terms of renormalization, we might call this a bare theory of complexity. For this theory, the behaviours of the UV (that is, short distances) and the IR (that is, long distances) are very different. The UV has violently large curvatures and a very short distance to the cut locus. Using the bare theory, computing complexity growth in the UV (short-distance behaviour) is straightforward. We find a linear growth with a very large slope. However, the calculation breaks down once the geodesic we are following passes the cut locus, in which non-perturbative effects become important. These effects slow the growth of complexity, and if our conjectures are correct, eventually the complexity growth becomes linear again, but with a much-reduced slope. A new schedule of penalty factors—the critical schedule—defines an effective theory that is easy to use in the IR.

In statistical mechanics and quantum field theory, this is analogous to the statement that a field theory is a flow between a UV conformal field theory and an IR conformal field theory. This means (among other things) that certain correlation functions in field theories exhibit a power-law decay in the UV and a power-law decay in the IR, but with different (anomalous) logarithmic slopes (referred to as critical exponents) in the UV and IR. Here the slopes of the linear growth of distances play the part of the logarithmic slopes in statistical physics. Our conjecture that the IR slopes differ dramatically from the UV slopes is analogous to the statement that in a strongly coupled field theory, anomalous dimensions are typically large.

The values of the penalty factors $\mathcal{I}_k$ are the parameters of the theory, playing the role of the set of (inverse) coupling constants in a quantum field theory. If a given penalty factor is greater than the value it attains in the critical schedule, then that parameter is irrelevant—that is, perturbing it does not affect the IR behaviour. The penalty factor becomes relevant only when it has the same value it would have had on the critical schedule, and any further decrease in $\mathcal{I}_k$ beyond this point then changes the distance function in the IR.

In describing the geometry of the group manifolds, we have used the terms of Wilsonian quantum field theory: UV and IR, bare theory, anomalous dimension, non-perturbative, effective theory, flow, coupling constants, relevant and irrelevant. At the moment the similarities between complexity geometry and the renormalization of quantum field theories are far from a precise isomorphism, but they are suggestive of deeper connections.

### Connection to coarse geometry

We explain the relationship of our work to the mathematical subject of coarse geometry and geometric group theory[42,43]. Supplementary Information 1 further rephrases our investigation and conjectures in this language, but the equivalences discussed there are somewhat less coarse than those allowed under the standard definitions reviewed here.

The main idea of coarse geometry is that given two metric spaces $X, X'$ equipped with distance functions $d, d'$ we can say that they are coarse equivalent or quasi-isometric $d \sim d'$ iff there exists a map $f: X \to X'$ such that

$$c^{-1}d(x_1, x_2) - a \le d'(f(x_1), f(x_2)) \le cd(x_1, x_2) + a \qquad (21)$$

for some $c \ge 1$ and $a \ge 0$. Furthermore, it is required that every point $x' \in X'$ is at most a fixed distance $b \ge 0$ from some image point $f(x)$, where $x$ could depend on $x'$. For our purposes, we apply this definition to the same underlying space $X = X'$ equipped with two different distance functions and take $f$ to be the identity. We then say that the two metrics are coarse equivalent $d \sim d'$ iff there exist $a$ and $c$ such that

$$c^{-1}d(x_1, x_2) - a \le d'(x_1, x_2) \le cd(x_1, x_2) + a. \qquad (22)$$

For an unbounded metric space, the statement has content because $a$ and $c$ are required to be finite. For a bounded space such as a metric on a finite group or a compact Lie group, the statement has no content unless we upper-bound $a$ and $c$. In the context of complexity geometry, it is natural to consider a sequence of metric spaces $X_n$, for example, $X_n = \text{U}(2^N)$. Then we would say that the sequences of geometries are coarse equivalent if we can find some constants $a$ and $c$ that are independent of $n$. If the diameter of $X_n$ is unbounded as $n \to \infty$ this is a non-trivial statement.

The notion of coarse equivalent or quasi-isometry defines an equivalence relation on the set of metrics $d$ on a given space. In our context, we are interested in the case in which the space is a Lie group $G$. In general, we can fully specify a left-invariant geometry on a Lie group of dimension $\dim(G)$ by specifying a $\dim(G) \times \dim(G)$ matrix $\mathcal{I}$ worth of parameters (which we refer to as penalty factors), which can be viewed as the infinitesimal line element near the identity element of the Lie group[2]. Hence in this context, we are discussing equivalence relations on these penalty factors $\mathcal{I} \sim \mathcal{I}'$. As mentioned above, this equivalence relation is meaningful as stated for a non-compact Lie group in which the diameter of the geometry is infinite with any reasonable choice of penalty factors. For sequences of compact Lie groups $X_n$, for example, $\text{U}(2^N)$, we consider corresponding sequences of penalty factors $\{\mathcal{I}_n\}$ and define an equivalence relation between such sequences $\{\mathcal{I}_n\} \sim \{\mathcal{I}'_n\}$.

More generally, we can imagine adjusting this criterion in various directions. For example, we could require that $a$ and $c$ are not independent of $n$ but have a mild $n$ dependence. In the Supplementary Information, we mention a bound[44] that was proved in the context of nilpotent Lie groups of the form

$$|d(x_1, x_2) - d'(x_1, x_2)| \le O(d(x_1, x_2)^{\alpha}) + a \qquad (23)$$

where $0 < \alpha < 1$. This implies that the fractional error vanishes at large $d$ at a rate no slower than $d^{\alpha-1}$. This is a stronger statement than equation (22) if $c$ is left unspecified, but it is a slightly weaker statement than equation (22) if it requires $c = 1$.

In the context of discrete groups, an elementary result is that the coarse geometry defined by the Cayley graph is independent of the choice of generating set[43]. That is, we can consider a group $G$ that is generated by some set of easy elements $g_1, \ldots, g_k$. The distance from the identity to some group element $g$ is the minimum length of the word formed from $g_1, \ldots, g_k$ that expresses $g$. Although the metric depends on the choice of generating set (or more generally, the penalty factors associated with each group element), the claim is that different choices of generating sets give distance functions that satisfy equation (22). Furthermore, we can consider properties of the geometry that depend on only the equivalence class. A particularly interesting property is $\delta$-hyperbolicity, which is a notion of negative curvature that applies even in this discrete context. We can identify negative curvature by observing that all triangles in negatively curved spaces are slim—that is, any point on one side of the triangle is close to some point on another side of the triangle, with the maximum separation set by the curvature scale. This property defines what is known as Gromov hyperbolic groups[45] and is the subject of ongoing mathematical work. A simple example is a free group, where the Cayley graph is an infinite tree. This notion of negative curvature may explain our conjecture that the critical metric has negative sectional curvatures[7,38]. In particular, the notion of $\delta$-hyperbolicity shows that the concept of large-scale curvature is not a contradiction. In the context of Lie groups, we expect that although many members of a given equivalence class exhibit extreme local curvatures, their large-scale curvatures (for example, that probed by large triangles) should approximately agree with the large-scale curvatures of the critical metric.

This work calls for an extension of the geometric group theory program to cover non-compact Lie groups and sequences of compact Lie groups. Furthermore, in analogy to the Cayley graph and discrete groups, we believe that in many cases, the number of equivalence classes of coarse geometries is small, despite there being a naively infinite number of different right-invariant metrics on Lie groups.

**Acknowledgements** M.H.F. acknowledges the Aspen Center for Physics for their hospitality in summer 2019 and 2021. L.S. is supported in part by the NSF grant PHY-1720397.

**Author contributions** All authors contributed to the paper and extensively discussed the results.

**Competing interests** The authors declare no competing interests.

**Additional information**

**Correspondence and requests for materials** should be addressed to Adam R. Brown.

**Peer review information** *Nature* thanks Fernando Pastawski and the other, anonymous, reviewer(s) for their contribution to the peer review of this work.

**Supplementary information**

# Universality in long-distance geometry and quantum complexity

In the format provided by the authors and unedited

## Supplementary Information for
# Universality in Long-Distance Geometry and Quantum Complexity

Adam R. Brown, Michael H. Freedman, Henry W. Lin, Leonard Susskind

# 1 Zooming in and zooming out

In this appendix, building on the Connection to Coarse Geometry section in the Methods, we describe a mathematical program to characterize the equivalence classes of metrics on right-invariant Lie groups under a certain equivalence relation. The equivalence relation should be loosely thought of as "zooming out" and only seeing the "coarse" features of the geometry. A possible definition will be presented in Eq. S.2, but first we review some related ideas in the mathematics literature.

As long as Riemannian manifolds have been studied,[1] the idea of fixing a point $x$ and zooming in, that is, multiplying all distances in the ball of radius $r$, $B_r$, by $1/r$, $r \ll 1$, has been central. The limit of such rescaling is the tangent space $T_x$ at $x$, a linear space with a metric.[2] In this paper we consider families of metrics where certain tangent directions are increasingly penalized, that is, are declared longer and longer. In the limit where motion in some direction is not just penalized but forbidden, one arrives at *sub-Riemannian* geometry. A fruitful observation of Gromov [1] is that zooming into a point on a sub-Riemannian manifold[3] endows the tangent space with the structure of a *self-similar nilpotent Lie group*. (The simplest example is the Heisenberg group, which has $[X, Y] = Z$, $[X, Z] = 0 = [Y, Z]$, with $Z$ rescaling as $l^2$ when $X$ and $Y$ are rescaled[4] by $l$.) On the other hand we may instead zoom out by considering $r \gg 1$. Such limits[5] of zooming

[1]Going back to the 19th century work of Gauss and Riemann.

[2]That is, a nonsingular inner product.

[3]The finite norm directions are assumed to generate all of $T_x$ under bracket, and an additional uniformity condition is assumed, which is always satisfied by right-invariant metrics on a Lie group.

[4]More generally, the Ball-Box theorem describes how $\epsilon$-balls in self-similar Lie groups are approximated by Riemannian boxes with $O(\epsilon^{k_i})$ side lengths where the integers $\{k_i\}$ are governed by commutator depth.

[5]In the physics literature, this is sometimes referred to as Wigner group contraction [2]. For example,

out are called *asymptotic cones*. Simply connected nilpotent Lie groups $G$ and their co-compact lattices are well-studied in this regard. The details of exactly how quickly the asymptote is approached, and how its geometry compares with the original Lie group geometry, has been an active area of research for 25 years (see [3]).[6]

We found, at least for the low-dimensional cases we could solve exactly, that the computed convergence is actually much faster than the proven estimates[7] (Prop. 3 and Thm. 4 of [3]). We find only a small[8] additive error separates the approximate from the limit. We do not know the scope in which such additive estimates hold, but they have inspired us to believe that in the case of interest to us, $\mathrm{U}(2^N)$, the large-scale behavior of at least a large class of right-invariant metrics may be characterized by broad phases with nearly constant large-scale geometry separated by sharp phase transitions. Let us at least describe a conjectural picture here, and also its conjectural relation to renormalization.

Before describing the picture we need to address the compactness of $\mathrm{U}(2^N)$. Zooming in is always possible. Traditionally, zooming out (choose $r \gg 1$) is restricted to non-compact spaces. This may be summarized by saying "all compact spaces have the coarse geometry of a point". The most famous of all "zoom out" theorems is Gromov's Theorem [4,5] that a finitely generated group of polynomial growth contains a nilpotent subgroup of finite index. Remarkably, this theorem has a beautiful quantitative version [6] applicable to *finite groups*[9]. Since it will be established in [8] that $\mathrm{U}(2^N)$, for many penalty schedules, has diameter exponential in $N$, the finitary

---

SE(2) is a Wigner contraction of SU(2): keep $\mathcal{I}_x = 1$, and take $\mathcal{I}_y$ & $\mathcal{I}_z$ infinite at fixed $\mathcal{I} \equiv \mathcal{I}_z/\mathcal{I}_y$.

[6] To understand zooming in and out the following pair of examples is instructive: If we start with the Euclidean plane both the zoom in and zoom out are again a Euclidean plane (the tangent space at a point and the asymptotic cone at that point respectively). If we start with the Hyperbolic plane and zoom in to a point we get its tangent space (as expected), if we zoom out we get what is called an R-tree, and infinitely branching dendritic object of topological dimension 1 which branches at a dense subset.

[7] The context of these estimates is again nilpotent Lie groups, but its relevance to metrics on $\mathrm{U}(2^N)$ is discussed in the next paragraph.

[8] For the SU(2) example with a penalty $\mathcal{I}$, Eq. 4 of the main text found that the additive error is $\mathrm{O}(\mathcal{I}^{-1/2})$.

[9] See also [7] for some discussion of coarse geometry in the context of a finite group model of complexity geometry.

philosophy (essentially controlled rescaling) can be applied to think about intermediate- and large-scale geometry of this very large, albeit compact, space.

We interpret the distance function on $\mathrm{U}(2^N)$, which depends on a choice of right-invariant[10] metric $g$ and amounts to defining the distance from $\mathbb{1}$, as the complexity. Equation 10 from the Methods is equivalent to the statement that this metric is given by

$$\mathrm{dist}_g(U) = \mathrm{inf}_\gamma \int_0^1 ds \langle \dot{\gamma}, \dot{\gamma} \rangle_g^{1/2}, \tag{S.1}$$

where the infimum is over piece-wise smooth paths from $\mathbb{1}$ to $U$. The same formula applies to the sub-Riemannian case, $\mathcal{I}_k \to \infty$. (We remind the reader that $\mathcal{I}_k$ are the diagonal elements of the metric on Pauli words of weight $k$, see Eq. 10 of the Methods.)

Since zooming out tends to blow up curvatures, even if one starts with a Riemannian metric, one may expect a sub-Riemannian (or worse) asymptotic cone in the zoom-out limit. Returning to the Heisenberg group, rescaling a right invariant metric, will, in the limit result in a Carathéodory geometry. Slightly more generally, much has been understood regarding scaling limits of nilpotent Lie groups, [3]: 1. How these rescaled spaces approach their limits, (see Thm 2 of [3]), 2. What is required for two different geometries to have the same limit, (see Prop 3 of [3]), and 3. How finitely generated subgroups approximate the continuous geometry (see Thm 4 of [3]). We feel all three discussions are relevant to our program of coarsely classifying right-invariant metrics on $\mathrm{U}(2^N)$.

In the cited results, the general pattern is that large-scale behavior is controlled to some tolerance, a power law intermediate between an constant additive and constant multiplicative error, by the behavior of the metric when reduced to the Abelianization $\mathfrak{g}/[\mathfrak{g}, \mathfrak{g}]$ of the group's Lie algebra.

The analogy we see is that our "easy" directions (in $\mathfrak{u}(2^N)$), being bracket complete, play the role of $\mathfrak{g}/[\mathfrak{g}, \mathfrak{g}]$. We expect metrics which agree on easy directions will have very similar intermediate and large-scale properties, corresponding to 2 above. Points 1 and 3 also play a role in the analogy. Comparison of Riemannian spaces to their sub-Riemannian limits should be a proof technique for extending 2 from nilpotent Lie groups to $\mathrm{U}(2^N)$. But the translation will require ideas such as in [41] because $\mathrm{U}(2^N)$ is compact, so the "approach to infinity" must be carefully quantified. Point 3 should help us make connections between the gate-based, and differential-geometric methods of navigating $\mathrm{U}(2^N)$.

[10]In the mathematics literature it is common to talk about left-invariant (but not necessarily right-invariant) metrics, but here we follow the (mirror-image) physics convention of considering right-invariant (but not necessarily left-invariant) metrics.

Here is a goal for making the analogy concrete. Give a criterion capable, in interesting cases, of splitting the spectrum to identifying the "easy" subspace $E \subset \mathfrak{u}(2^N)$. Then $g_0$ is defined to be the sub-Riemannian metric obtained by simply retaining the metric on $E$ and forbidding motion in other directions. If $E$ is *effectively* bracket complete (effective in the sense that the completion is achieved with reasonable numerical efficiency), then we would like to establish the following conjectural inequality:

$$|\mathrm{dist}_g(U_1, U_2) - \mathrm{dist}_{g_0}(U_1, U_2)| \leq \mathrm{O}((\mathrm{dist}_g(U_1, U_2))^{1-\alpha}) + C_{g_0} \qquad \text{(S.2)}$$

for $U_1, U_2 \in \mathrm{U}(2^N)$, some $\alpha \in (0, 1)$, and $C_{g_0} > 0$ is a constant which is order $O(1)$ when $E$ efficiently generates $\mathfrak{g}$ under brackets.

For large scales, comparable to the diameter, the most optimistic conjecture is that distances induced by the sub-Riemannian metric $g_0$ are *additively* close to those of an approximating Riemannian metric $g$. We require, of course, that $g$ agree with $g_0$ on the bracket generating subspace $E$ (for example 1- and 2-qubit Pauli words) and that the penalties defining $g$ in orthogonal directions are reasonably large.[11] We do not have in hand technical tools powerful enough to attempt a proof. Rather than sketching a proof, contrariwise, we will explain a few fundamental obstacles to finding one.

The perturbative calculations in [9] seem to have reached their natural limit and fresh ideas are needed. As discussed in the Methods, in [8] it will be shown that metrics with a sufficiently rapidly growing penalty schedule have diameters exponential in $N$. It will further be shown [9] that a $U$ with distance $d$ from $\mathbb{1}$ in a sufficiently expensive penalty metric can be $\epsilon$-approached by a path of length $\epsilon^{-1} d^2$ in the sub-Riemannian cliff metric, see Eq. 18 of the main text. Here "$\epsilon$-approached" means the path arrives within a Killing-metric ball of radius $\epsilon$. However, the ball-box theorem warns us that $\epsilon$-approach may still leave us a tiny power of $\epsilon$, like $\epsilon^{0.001}$, away in the metrics of interest. This phenomenon is well-known for nilpotent groups[12] [3] where it is captured by the notion of *abnormal geodesic arc*. This is a geodesic arc in the sub-Riemannian geometry whose $\epsilon$-variations fail to cover any Riemannian $O(\epsilon)$ balls around its endpoints. So, notoriously, it is hard to solve your problems at the last moment by first arriving Killing-close and then proceeding to the desired target. This has led us to ask if there is a *lever arm* here to exploit: If we recognize, far off, that we will slightly (in the Killing norm) miss our target, can we make inexpensive adjustments early (a strategy we called 'wiggle room')? If possible, learning how to do this seems tantamount to proving at least a weakened version of our conjecture, where even beyond a multiplicative constant we would allow a power-law expansion of distance.

[11] This is the sense in which $g$ *approximates* $g_0$, where the normal penalties are infinite.

Actually, Ref. [10] provides further evidence that the critical/cliff correspondence pertains not only at the largest scale (the diameter) but also at intermediate scales. This is addressed by comparing covering numbers $c_{\text{critical}}(r)$ and $c_{\text{cliff}}(r)$, the number of $r$-balls required to cover the manifold. For $r$ = diameter, this number is 1; as $r \to 0$, it is a surrogate for volume. As $r$ varies, the covering number probes the geometry on a range of scales (see the discussion of marbles in the main text[13]).

So for the case where only 2-qubit interactions are easy, our conjectural picture is that there is some *critical* metric $g_{\text{crit}}$ determined by a critical schedule of penalties $\bar{\mathcal{I}}_k$, with $\mathcal{I}_1$ and $\mathcal{I}_2$ fixed, which lies at the *lower* edge of the phase of metrics which at its upper limit is the metric with $\mathcal{I}_{k\geq 3} = \infty$, the sub-Riemannian metric. Thus little should happen to already large distances when penalties $\mathcal{I}_{k\geq 3}$ are increased but if any $\mathcal{I}_k$ are reduced certain large distances shrink at first order.

It is interesting to ask: is $\bar{\mathcal{I}}_k$ unique (or perhaps forming a convex set?). If it is unique, perhaps it admits several other characterizations. Also, what kind of renormalization group flow might take us toward $\bar{\mathcal{I}}_k$?

[12] Nilpotent groups are relevant to us both in the small where they arise [6] as a structure on the tangent space of a sub-Riemannian manifold, and conjecturally in the zoom-out limit, provided approximate asymptotic cones can be defined.

[13] Since in flat $T^d$ the number of marbles of radius $n$ we can cram in scales like $n(r) \sim r^{-d}$, the $n(r) \sim r^{-4}$ scaling we saw at some scales for the 'three-dimensional' Berger sphere is characteristic of four dimensions (see e.g. [11]), and so exhibits what in quantum field theory is called an 'anomalous dimension'.

Among metrics with a prescribed fixed value on 1- and 2-qubit Hamiltonians, let us speculate on possible characterizations of $\bar{\mathcal{I}}_k$. A first possibility is that using the natural partial order on norms, $\bar{\mathcal{I}}_k$ should be an infimum penalty schedule among those where the first variation of diameter vanishes, $\frac{\partial(\text{diameter})}{\partial g} = 0$ (when the variation is presumed constant on the space spanned by 1- and 2-qubit Hamiltonians), and a supremum metric among all such metrics with $\frac{\partial(\text{diameter})}{\partial g} \neq 0$. Another possibility is to use the spectrum of the Laplacian, e.g., $\lambda_1$ the first non-zero eigenvalue of $\nabla^2$ on the group manifold. Motivated by [12] where overcompleteness drives up a lowerbound for $\lambda_1$, the critical metric (from a very broad class—perhaps even including Finsler metrics, e.g. taxicab metrics) may be the one maximizing $\lambda_1(\text{diameter})^2$, again constrained on 1- and 2-qubit Hamiltonians. Furthermore, the collapse of sectional curvatures, $|K|$, in the qubit context near the (perhaps critical) exponential metric can be seen as reinforcing this conjecture. For example, among all geometries on the 2-sphere with diameter $\pi$, $\lambda_1$ is apparently maximized by the round metric. Relatedly, a probe that is sensitive to parallelism could come from the notion of conductance, if we imagine that the complexity geometry is fabricated from a material of constant resistivity. Perhaps the critical metric maximizes $c(\text{diameter})^{2-4^N}$, where $c$ is the average conductivity between all pairs of points $U_1, U_2 \in \mathrm{U}(2^N)$, constrained as above. This relates to the 'load-balancing' property we expect for the critical metric. A final possibility is that the critical metric could minimize the average distance to the cut locus over the visual sphere, constrained as above.

To complete our conjectural picture, imagine a flow which seeks to lower all $\mathcal{I}_k$, $k \geq 3$, but constantly and homothetically rescales $\{\mathcal{I}_3, \mathcal{I}_4, \dots\}$ to maintain constant diameter. A flow of this type, if initialized in the stable phase near the sub-Riemannian $g_0$, would be expected to converge to $g_{\text{crit}}$. One can generalize this flow to general right-invariant Riemannian metrics (again fixed on the span $L$ of 1- and 2-qubit Hamiltonians) to a *random* flow. Randomly select an eigen (principal) direction of $g$ in $L^\perp$, shorten it, and then homothetically rescale $g|_{L^\perp}$ to keep the diameter constant. A flow of this type might be considered as a kind of renormalization of right-invariant metrics on $\mathrm{U}(2^N)$. A natural question is whether the flow of geometries from 'UV' to 'IR' has anything to do with Ricci flow, which has a known connection to the renormalization group of 2D quantum field theories. (See e.g., [13] for a discussion of Ricci flow in the context of homogeneous geometries on group manifolds.) This seems unlikely to us as the conjectured "fixed points" are not Einstein metrics, although perhaps there is a generalization of the concept that could be applicable.

There is also the question: should the flow be defined only on metrics which know about some fixed qubit structure—as for those we have so far discussed—or in the spirit of [14] and [15] should we consider all right-invariant metrics and ask that the flow have fixed points which locate a qubit structure and a critical schedule together.

The large-scale geometry of right-invariant metrics, even for exceedingly simple penalty schedules, is enormously subtle. The successful proofs (e.g. diameter lowerbounds) are all indirect. To highlight how little is known explicitly, without complexity assumptions, there seems to be no explicit unitaries whose sub-Riemannian cliff metric distance from $\mathbb{1}$ is known to grow superlinearly with $N$; see however [16] for an interesting construction that relies on a complexity assumption about PSPACE.

Another question whose answer is not yet known: at long distances (comparable to the diameter) is there a "high frequency" component to $|\text{dist}_{\text{cliff}} - \text{dist}_{\text{crit}}|$? Essentially this asks if $d_{\text{cliff}}(\mathbb{1}, U) = \ell \sim$ diam and $d_{\text{crit}}(U, U') = \epsilon > 0$ then is $d_{\text{cliff}}(\mathbb{1}, U') \leq \ell + \mathrm{O}(\epsilon)$? A version of this question was addressed in the context of a specific 5-dimensional nilpotent Lie group [3]. The authors [3] showed that the "cliff metric" on this group has an abnormal geodesic segment, i.e. a geodesic arc of any desired length in some right-invariant sub-Riemannian geometry, whose $\epsilon$ variations fail to cover an $\mathrm{O}(\epsilon)$-Riemannian neighborhood of its endpoints. So in that example we would say there is a high-frequency discrepancy at all scales. On $\mathrm{U}(2^N)$ we know of no analogous abnormality and would guess that the very rich dynamics of geodesics should rule it out, but have no proof. Perhaps the simplest reason for expecting a smooth relationship is the lack of any candidate "quantum number."

More generally, beyond the complexity geometry examples on $\mathrm{U}(2^N)$, we may ask for a classification of the equivalence classes of coarse geometries of metrics on non-compact Lie groups, or sequences of compact Lie groups. Such a mathematical program would significantly generalize existing work on the coarse geometry of finitely-generated discrete groups.

# 2 Holography, load balancing, and the critical metric

In the Methods we described how the conjectures of this paper, if correct, could give a robust theoretical underpinning to the holographic complexity conjectures. In this section, we will describe further implications.

First, let's note that as well as being informed by the conjectures in this paper, the holographic complexity conjecture also in turn informs our expectations for the 'critical metric', defined in the Main Conjectures section of the main text. In the Methods, we described evidence that the binomial metric is in the same universality class as the cliff metric. Could the binomial metric be not only *in* the universality class, but be the cheapest member of the class, i.e., be the critical metric? On the one hand, the binomial metric shares with the exponential metric the property that the sectional curvatures are modest, which we argued is one signature of being the critical metric. On the other hand, if the binomial metric really were the critical metric, an odd consequence would follow from the fact the binomial metric has an $N$-dependence that changes with $k$, $\mathcal{I}_k \sim N^{2k}$. If the binomial metric were the critical metric, this would imply that the complexification rate of $k$-local Hamiltonians is never extensive in $N$ for all values of $k$, even if we multiply the entire penalty schedule by an overall $N$-dependent factor. This is different from the expected complexification rate of Brownian circuits, as well as the gate complexification rate, and from the rate we expect of black holes. This is one reason to prefer the exponential metric, given by Eq. 12 of the main text, as a candidate for the critical metric: for the exponential metric, $\mathcal{I}_k$ depends only on $k$ and is independent of $N$. (Though note also that for generic $k$-local Hamiltonians with $\text{Tr}[H^2] = 1$ the rate of change of *inner-product* distance is not extensive and instead scales like $\sqrt{N}$, essentially due to Pythagoras' theorem.)

Second, as a subject for further study, recall that we described how the critical metric is approximately load-balanced, meaning that there are many different paths, all with roughly the same complexity, that connect any two distant states. It would be interesting to explore what this degeneracy might mean in the context of bulk physics.

Finally, let's note that one should not confuse the sensitivity of the complexity to the choice of gates or penalty factors with the sensitivity of the complexity to the choice of UV regulator of the quantum field theory, discussed in [17–20]. For example, the SYK model is UV finite, but one still needs to discuss the choice of penalty factors in the definition of complexity.

# 3 Two-dimensional example

In the introduction of the main text, we observed the surprising fact that a deformation of the metric that has only a tiny additive effect on the distance between any pair of points can nevertheless drive the volume to infinity. We illustrated this with a three-dimensional homogeneous metric, the Berger sphere. Let's now give an example of this phenomenon in two dimensions. The advantage of the two-dimensional example is that it is easier to visualize; the disadvantage is that we must consider a metric that is inhomogeneous.

The two-dimensional example is the metric described in Sec. 3 of Ref. [21]. While our three-dimensional example in the main text was the *unitary* complexity of a single qubit, given by a deformed three-sphere (a non-bi-invariant metric on SU(2)), our two-dimensional example is the *state* complexity of a single qubit, given by a deformed two-sphere (a deformed Bloch sphere). The distance between two points on the deformed Bloch sphere is defined as the *unitary* complexity of the least complex unitary that connects them, where the unitary complexity is given by Eq. 1 of the main text. This is explained in detail in Ref. [21], along with a geometric visualization. The line element is

$$4\, ds^2 = d\theta^2 + \frac{\mathcal{I} \sin^2\theta}{\mathcal{I}\cos^2\theta + \sin^2\theta} d\phi^2. \tag{S.3}$$

For $\mathcal{I} = 1$ two points are separated by the standard angular distance on a round two-sphere (the inner-product distance). For $\mathcal{I} > 1$ rotations of the sphere are more expensive when the axis of rotation has a component in the vertical direction, $\sigma_z$. As $\mathcal{I}$ becomes ginormous, the only rotations that are not prohibitively expensive have rotation axes that are equatorial.

To manifest the surprising phenomenon, we start with $\mathcal{I}$ very large, and then take the limit $\mathcal{I} \to \infty$. On the one hand this makes only a tiny additive change to distances, since (via its definition) state complexity inherits Eq. 4 from unitary complexity. On the other hand the volume goes (logarithmically) to infinity, since two neighbouring points on the equator have their distance multiplied by a huge factor, $\sqrt{\mathcal{I}}$. The limit of the geometry as $\mathcal{I} \to \infty$, and its relation to the optimal shape for tunnels bored through the Earth, is described in detail in Ref. [21].